\documentclass[review]{elsarticle}

\usepackage{hyperref}

\journal{Journal of \LaTeX\ Templates}

\begin{document}

\begin{frontmatter}
\title{Testing a new model of local wind erosion and dust deposition on field campaign data in Morocco. Adaptation of the model for Mars.}

\author[oac]{N. V. Deniskina\corref{corauthor}}
\cortext[corauthor]{Corresponding author: N.V.Deniskina}
\ead{nvdeniskina@gmail.com}


\address[oac]{Capodimonte Astronomical Observatory, Naples, Italy.}

\begin{abstract}
Global and mesoscale models represent the background (slowly varying) winds on Mars, but short timescale wind variability is not explicitly represented. The local wind erosion and dust deposition model can be useful for more accurate local wind friction, wind friction threshold, horizontal sand flux and vertical dust flux modeling on Mars. 
Such a model based on the model of Zender \cite{Zender} is proposed in this article.
It includes different methods of threshold wind friction estimation and takes into account the influence of atmospheric electric fields on sand and dust elevation processes.
 The model parameterized for the Earth was tested on data acquired in Morocco field campaign 2013-2014 (Esposito et al.,) \cite{Esposito}, \cite{Esposito2}. Wind friction, wind friction threshold, horizontal sand flux and vertical dust flux were simulated by using the following inputs measured in the field: wind speed, wind direction, air temperature, air pressure, air humidity, solar irradiation and surface roughness length. The same inputs are supposed to be measured by DREAMs equipment on Mars (except for the surface  roughness length which can be estimated for Mars according the method of Hebrard \cite{Hebrard}). 
Simulated and experimental values have shown a good agreement.  
 The model was adapted to Martian conditions and applied to the data of Viking lander. 
The local wind erosion and dust deposition model parameterized for Mars may serve 1) to analyze local lander/rovers data, 2) a source of local parameters/inputs (like threshold wind friction velocity,
sand and dust fluxes) for the MGCM (Mars Global Circulation Model). 

\end{abstract}
\begin{keyword}
\texttt{Wind erosion and dust deposition model, MGCM, wind friction, vertical dust flux}
\end{keyword}
\end{frontmatter}
\section{Introduction}
Haberle \cite{Haberle} first used a model to show the dynamical response to a dusty Martian atmosphere. 
Now there are global and mesoscale models that provide predictions of atmosphere condition changes like slope wind and dust storms on Mars.  
Detailed description of the Mars climate modeling is presented in the works of Kahre et al. \cite{Kahre}, Spiga et al. \cite{Spiga}, Mulhholland et al. \cite{Mulholland}, \cite{Mulholland2}, 
Newman et al. \cite{Newman}. 
These models are  mathematical simulations of the general circulation of a planetary atmosphere.  
They use the Navier–Stokes equations on a rotating sphere with thermodynamic terms for various energy sources like radiation and latent heat. The dynamical core of these models 
integrates the equations of fluid motion for surface pressure horizontal components of velocity in layers, temperature and water vapor in layers and radiation. 
There are prognostic equations that are a function of time (typically for winds, temperature, moisture, and surface pressure) 
together with diagnostic equations that are evaluated from them for a specific time period.
These equations are a basis for computer programs which simulate the Martian atmosphere and are used for weather and climate forecasting.   
Meteorological and surface parameters measured experimentally or simulated are needed for models functioning.   
Data of Mars orbital and lander  missions are used as inputs for global/mesoscale simulations to retrieve unknown parameters that cannot be measured experimentally. 
This procedure is applied to global and mesoscale models because grid cells are big and it is possible to use parameters averaged over a large grid cell. The outputs of the modeling can be inserted in the code to retrieve the meteorological parameters and 
compare them with measurements of satellites/rovers in order to test simulations and adjust parameters. Dust elevation physics is simplified for 
global and mesoscale simulations, thus a smaller number of parameters is required.    
 
Mars climate modeling give promising results Kahre et al., \cite{Kahre},  Spiga et al., \cite{Spiga}, Mulholland \cite{Mulholland}, \cite{Mulholland2}, Newman et al., \cite{Newman}, but improvements are still needed. 
For example, annual variability of dust  movements on Mars (dust storm size, timing and location of origin) has not been adequately captured by
Mars climate modeling.
Some degree of atmospheric variability leading to variation in storm magnitude from year to year can be simulated by MGCM if the wind
stress lifting parametrization uses a sufficiently high threshold (Basu et al., 2006 \cite{Basu}). 
The problem is that this threshold value is too high  to permit dust storm on Mars.  
A more accurate atmospheric dust input (local wind input) for MGCM can possibly resolve the discrepancy. 

Gierasch and Goody first identified that dust could profoundly influence the thermal structure of the Martian atmosphere\cite{GG}. Atmospheric dust is a crucial component of the Martian climate system, and is a major driver of much of the inter annual variability observed in atmospheric
 circulation and climate on a range of timescales according to Liu et al., 2003, \cite{Liu}, Basu et al., 2006 \cite{Basu},  Mulholland \cite{Mulholland},
Rotstayn et al., \cite{Rotstayn}, Shao et al., \cite{Shao2}, Fenton et al., \cite{Fenton}, \cite{Fenton2}, Mahowald et al., \cite{Mahowald}.
The  precise parametrization of dust emission processes  is very important (Newman et al., 2002b  \cite{Newman}; Basuet al., 2006  \cite{Basu}; 
Kahre et al., 2008 \cite{Kahre}). 

The global and mesoscale Mars climate simulations compared with local wind erosion simulation is lacked the detailed representation of this process. It is not easy  to  represent dust lifting parametrizations for Mars because of the fact that lifting thresholds and vertical mass 
fluxes were never measured directly on Mars and there are many vitally important factors for theoretically determining these quantities (interparticle cohesion, surface size distribution and
composition, etc) that are not well explored on Mars.  For this reason, simplifying assumptions such as constant stress thresholds for lifting have been typically  employed.

Besides, the Mars Global Circulation Model as well as most general global and regional circulation models, does not resolve short-lived, small scale waves, eddies or convective events, but rather produces smoothly varying
balanced fields, representative of the more slowly varying large- to global-scale components of the atmospheric circulation. 
MGCM wind velocities, typically output every half an hour, may thus be thought of as representing the background, slowly varying winds on Mars, but short timescale variability is not explicitly represented by the model. 
Actual winds on Mars are distributed about such representative values, depending on the degree of gustiness  \cite{Newman}. 
So even if the MGCM is able to correctly model the background wind on  
Mars, a simple parametrization like "if wind friction speed $>$ threshold wind friction speed" will fail to capture dust lifting by strong gusts 
during a period when the mean wind is below threshold \cite{Newman}. 
In order to combat this problem the wind distribution may be modeled using a Weibull distribution, such as is used in the terrestrial field of 
renewable energy (e.g.,Zender \cite{Zender}, Seguro and Lambert, 2000 \cite{Seguro}).
Lorenz et al., 1995, 1996 \cite{Lorenz}, \cite{Lorenz2}, used Viking Lander hourly averaged wind speed data, for periods of a few sols at a time, 
to derive the best fit Weibull distributions. They obtained estimates of representative wind over a few sols and the gustiness parameter, 
which accounts for variations on hourly timescales \cite{Newman} and have shown that the parameterization for global/regional modeling, however, 
requires a gustiness parameter to account for variations on minute timescales. 

The works of Mulholland \cite{Mulholland2} introduce a similar approach to wind friction threshold estimation considering sub-grid scale
variability of both the near-surface wind field and the surface roughness and confirm that grid -scale parameterization will fail to capture dust lifting by strong gusts during a 
period when the mean winds are low.

Large-eddy simulations of Michaels and Rafkin proposed for Mars \cite{Michaels} resolve big-scale eddies (or convective events) and are applied for dust devils simulation and/or description of dust devil mechanism of 
dust lifting. These modeles are local but does not resolve short-lived, small scale wind waves responsible for lifting by near-surface  wind stress that supposed to be a principal mechanism of dust lifting on Mars \cite{Newman}, 
\cite{Mulholland}. 
 
 The local model of Kok et al., \cite{Kok} that simulates sand and dust fluxes on Mars uses simplifying assumptions for dust lifting process such as, for example,
 constant stress threshold. The code  needs  constant wind friction and constant wind friction 
threshold as inputs and uses very time consuming algorithm to approach to the saltation steady state. But an important point is that the model of Kok takes into account 
the electric field that can be an important factor for sand/dust lifting on Mars.

A local wind erosion model which permits to account for variations on small (seconds) timescale and simulate wind friction and wind friction threshold at each 
input data point is needed. Some time ago, the implementation of local model was impossible due to the lack of model parameters.
Now, the parametrization of the model can be supplied by combining meteorological and surface data of Martian rover and orbiter
missions and can be compared with global and regional Mars simulations results, for  example, by using Mars Climate Database v5.2. ($http://www-mars.lmd.jussieu.fr/mcd_{}python/$).
The surface roughness length can be estimated by the method of Hebrard \cite{Hebrard}.    

To sum up, at the moment, there is no detailed local wind erosion/dust deposition soft that would take into account short-term changes in horizontal wind, 
wind stress, wind friction threshold, stability of the atmosphere and the electric field on Mars.  

This paper  proposes a new local dust emission model  for detailed calculation of  wind friction, wind friction threshold as well as horizontal sand flux and vertical dust flux at  each time-step. 
The model simulates the distribution of  dust in the atmosphere,  mobilization by wind, dry and wet deposition and
transport processes. 
 
 The model 

1)combines the detailed description of the local dust emission process with possibility to include various algorithms of wind friction calculation;

2) includs atmospheric electric field influence on dust lifting process; 

3) includs accurate stability corrections developed for terrestrial models like in work of  Zender et al. \cite{Zender}, Skamarock and Klemp, 2008 \cite{Skamarock}. 

The model is based on the local model of Zender \cite{Zender} for the Earth and can be parameterized  as for Mars as for the Earth.
Dust lifting on Mars is represented as lifting by near-surface  wind stress like in 
Newman et al., 2002a \cite{Newman}, Mulholland \cite{Mulholland}, Kok et al. \cite{Kok}.

The model (parameterized for the Earth) was tested on the data obtained experimentally in Morocco desert \cite{Esposito}. 
The same parameters that supposed to be measured by DREAMS mission on Mars were  used as model inputs: wind speed and direction, air temperature, air pressure, air humidity, solar irradiation. The surface roughness length, that was measured experimentally in Morocco, can be estimated on Mars  by the method of Hebrard \cite{Hebrard}. 
 The outputs of the model are wind friction, wind friction threshold, horizontal sand flux and vertical dust flux.
All these variables (except vertical dust flux) were also measured experimentally in Morocco, and so we had an opportunity to verify the model.
 
The data obtained in the desert of Morocco were chosen for the test, since the climate and surface morphology there are similar to the Martian ones \cite{Esposito}.
The model showed a good fit to the experiment on all output parameters except the vertical sand flux that was not measured in Morocco and we had no opportunity to check it.  
Electric field can be important in the dust lifting process, especially on Mars. That is why we included in the algorithm the possibility to take into account the influence of the atmospheric electric field when calculating the output parameters. The base algorithm of Kok \cite{Kok} was used. 
We adapted the model for Mars conditions,  applied  it to Viking Lander data and made an estimation of  local wind friction threshold. 
At the moment the result can not be compared with the experimental value because wind friction threshold was not yet measured on Mars,
but it is near the value of wind friction threshold estimated by global and mesoscale simulations for Mars.  
\section{Model and Methods}
\subsection{Assumptions of the model}
  The following assumptions were used in the model:
 \begin{itemize}
	 \item The source process for dust is mobilization by wind. Thresholds for saltation, moisture inhibition, drag partitioning and saltation feedback are taken into account.
	 \item Soil texture is globally uniform and is replete with saltators. 
	 \item The microphysical and micrometeorological approach to dust mobilization developed by Marticorena and Bergametti, 1995 was used  \cite{Marticorena&Bergametti_95}.
	 \item Dust particles are not directly mobilized by the wind  but are primarily injected into the atmosphere during the sandblasting caused by saltation according to Alfaro and Gomes, 
	  2001 \cite{Alfaro} and  Grini et al., 2002 \cite{Grini}. Sandblasting refers to the disaggregation and ejection of clay by saltating sand-sized particles ($D_p>60 \mu m$).
	 \item Saltation initiates when the turbulent drag of the surface atmosphere dissipates enough momentum to overcome the gravitational inertia of sand-sized particles.
	\item  The kinematic and thermodynamic properties of the boundary layer are  determined by assuming that the surface and  and atmosphere constantly 
	adjust surface heat, vapor and momentum exchanges in order to maintain thermal equilibrium with the radiation field as described by Bonan, 1996 \cite{Bonan}.
	\item  The horizontal mass flux of saltating particles $Q_s$ is estimated according to the theory of White, 1979 \cite{White} and 
	depends on atmospheric density, wind friction speed, acceleration of gravity and wind friction threshold.  
	\item The wind friction threshold  in the formulation of $Q_s$ is the minimal wind friction threshold for the  particle of optimal size $D_0$.
	\item The horizontal sand mass flux $Q_s$ is converted to a vertical dust mass flux $F_d$ with  the sandblasting mass efficiency “$a$” ($F_d= aQ_s$ ) according to Alfaro and Gomes, 1997 \cite{Alfaro_1997}.  
	Size and drag-independent parametrization of "$a$" is adopted, “$a$” depends only on mass of clay.
	\item Saltation leads to dust production whenever $u* > u*_t(D_0)$. This assumption means that soils depleted in particles of size $D_0$  will begin saltation at unrealistically low wind friction speed.  
 \end{itemize}
 \subsection{ Inputs and outputs of the model}
The code permits to use different inputs and simulete more than 100 outputs parameters (like the model of Zender \cite{Zender}). 
The following inputs and outputs were used in this work.

Input (experimental data measured at one level of height):   
 \begin{itemize} 
 \item1) wind speeds and directions,
 \item2) air temperature T, 
 \item3) air pressure P ,
 \item4) solar radiation flux, 
 \item5) air humidity, 
 \item6) surface roughness length (measured in Morocco, estimated by the method of Hebrard for Mars). 
 \end{itemize}
 Despite the fact that the surface temperature was measured in an experiment in Morocco, an average  value  (taken from the global map distribution of surface temperatures )
was used for simulation.

Output (simulated data): 
\begin{itemize} 
 \item1) wind reference, 
 \item2) wind friction (estimated in three different ways),
 \item3) wind friction threshold, 
 \item4) horizontal sand flux,
 \item5) vertical dust flux.  
 \end{itemize}
\subsection{Wind reference estimation}
The reference wind was calculated from input wind value taking into account the logarithmic wind profile and applying stability corrections method of Monin- Obukhov as described by Bonan, 1996  \cite{Bonan}, 
 Large and Pond (1981,1982) \cite{Large1}, \cite{Large2} and also in model of Zender \cite{Zender}. 
 The model uses notation of Bonan, 1996 (page 50) \cite{Bonan} where the height of 10 m is reference height and second level is the height of the atmosphere.
\subsection{Wind friction $u^*$ estimation.}
Wind friction was estimated in three different ways: 1)by profile method; 2) by resistance method; 
3) by resistance method with Owen's effect correction.  
\subsubsection{Wind friction deposition ($u^*_{deposition}$). Profile method.}
$u^*_{deposition}$ was estimated according to the formulation of Bonan, 1996 \cite{Bonan} and Land and Surface model 
of Large and Pond (1981,1982) \cite{Large1}, \cite{Large2}. The applied stability correction is also described by Large and 
Pond (1981,1982) \cite{Large1}, \cite{Large2}. The profile method  is adopted for Global Circulation Model.
The iterative procedure is represented in Zender et al., 1991 \cite{Zender}. 
The routine uses the specified surface temperature  $T_{srf}$ rather than solving energy balance equation for new surface temperature on each step.
The formula was taken from  Jacovides et al., 1992 \cite{Jacovides}.  

\begin{equation}
\label{eq1}
u^*_{deposition}=
 =\delta u*ConstKarman/[log(H_2/H_1)-\psi_m(H_2/L)+\psi_m(H_1/L)] 
\end{equation}
 where:
 $L=T*u^{*2}/(g*ConstKarman*T_{scale})$ Monin Obukhov Length; 
 
 $T_{scl}=\delta T*ConstKarman/[R*[ln(H_{T2}/H_{T1})-\psi_h(H_{T2}/L)+ \psi_h(H_{T1}/L]]$; 

 $H_2$ and $H_1$ are the heights of wind speed measurements;
 $H_{T2}$  and $H_{T1}$ are the heights of air temperature measurements;
 
 $\delta u$  and $\delta T$ are wind and temperature differences between the respective levels;
 $T$  is the average surface temperature;
 $g$ is the acceleration due to the gravity;
 
$\psi_m$ and $\psi_h$ are empirical stability correction functions for momentum and heat described in Jacovides et al., 1992 \cite{Jacovides}. 

If the surface roughness length $z_0$ is given as an input, then the only one level of wind and temperature measurements is required:
 $H_2 = z_0$ and $u_2= 0$ in Eq. (1), $T2 \approx T_{surface}$ and the accuracy of the above procedure depends on the assumed value of $z_0$. 
 The Equation 1 can be represented as:
  \begin{equation}
\label{eq2}
 u^*_{deposition} =u_{bnd}*ConstKarman/[log(H/z_0)-\psi_m(z_0/L)+\psi_m(H/L)]
 \end{equation}
where:  $u_{bnd}$=max($u$,$u_{min}$);    $z_0$ is the surface roughness length.
 
\subsubsection{Wind friction mobilization ($u^*_{mobilization}$). Resistance method.}
 The method is optimized for dust source regions: dry, bare, uncovered land.  
 $u^*_{mobilization}$ is estimated by using aerodynamic resistance calculation. The theory and algorithms were developed by Bonan, 1996 \cite{Bonan}. 
Surface kinematic fluxes of momentum, latent heat and sensible heat are defined from Monin-Obukhov similarity theory applied to the surface/constant flux layer.
The theory states that when scaled appropriately, dimensionless mean horizontal wind, mean potential temperature and mean specific 
humidity profile gradients depend on unique function $H-d/L$, where $H$ is the height, $d$ is the displacement height, $L$ is 
the Monin-Obukhov length.

$u$ is defined to equal zero at the height of  $z_o + d$. The aerodynamic resistance is calculated between the 
atmosphere and the surface at the height of $z_o + d$.
 At each step the routine solves the energy balance equation to adjust the surface temperature $T_{srf}$ .
 \begin{equation}
\label{eq3}
u^*_{mobilization}=u_{bnd}*[R_{aerodynamic}*u_{bnd}]  
\end{equation}
 where $u_{bnd}$=max($u$,$u_{min}$);
  
$R_{aerodynamic}=max(1/(ConstKarman^2*u_{bnd}*[log((H-H_0)/z_0)-Corr]^2,1)$ -- the aerodynamic resistance of air layer between the surface and the height 
of measurement according to Jacovides et al., 1992 \cite{Jacovides}. 
                       
$Corr$ -- is the Monin-Obukhov stability correction function Jacovides et al., 1992 \cite{Jacovides}.  

$z_o$ was estimated according to the method of Hebrard \cite{Hebrard}.

$d$ was taken equal to zero. 

\subsubsection{Wind friction saltation ($u^*_{saltation}$).}
 \begin{equation}
\label{eq4}
 u^*_{saltation}=u^*_{mobilization}+u^*_{Owen-effect-correction}
\end{equation}

The Owen's effect refers to the positive feedback of saltation on wind friction. The
increase in wind friction speed due to saltation varies quadratically with the difference between the 10 m wind speed $u_{ref}$ 
and the threshold wind speed at 10 m $u_{ref_t}$:  

$u^*_{Owen's-effect-correction}=0.003 (\delta u_{ref})^2$ - according to Gillette et al., 1992 \cite{GMB}, where $\delta u_{ref}=u_{ref}-u_{ref_t}$. 

\subsection{ Threshold wind friction ($u_t^*$) estimation.}

The threshold of wind friction  ($u_t^*$) was estimated according to Large and Pond, 1981 \cite{Large1}. 
\begin{equation}
\label{eq5}
 u_t^*=\sqrt{F_{inter particle} \rho_{fct} Re_{opt}}/\sqrt{\rho}
\end{equation}
where $Re_{opt}$ is the Reynolds number for optimal particle  described in Marticorena and Bergametti,1995 \cite{Marticorena&Bergametti_95} and Zender et al., 1991 \cite{Zender}. 

 $\rho_{fct}=\rho_{saltators}gD_{opt}$ - is density ratio factor for saltation (Iversen and White, 1982) \cite{Iv}.
 
 $F_{inter particle}=1+6^{-07}/(\rho_{saltators}g D_{opt}^{2.5})$ Iversen and White, 1982 \cite{Iv}. 

Two corrections can be applied for wind friction threshold: 1) “drag partitioning” correction and 2)“moisture inhibition” correction and the formula 5 may be rewritten in the following way (Zender et al., 1991) \cite{Zender}: 

 ${u_t}^*_{corrected}=({u_t}^*) ({u_t}^*_{correction-roughness}) ({u_t}^*_{correction-moisture})$
 
“Drag partitioning" correction takes into account the efficiency with which drag is partitioned between erodible and non erodible soils
 and depends on two parameters: surface roughness length $z_0 (m)$  and "smooth" roughness  length $z_{0s} (m)$ (Marticorena and Bergametti,1995) \cite{Marticorena&Bergametti_95}. 
 This correction was applied for the data of Morocco.
“Moisture inhibition" correction is represented by the parametrization of Fecan et al., 1999 \cite{FMB} and is useful when the near-surface
soil gravimetric water content $w$ exceeds a threshold gravimetric water content $w_t$: $w > w_t$ or in other words when the 
capillary force suppress dust deflation.  It was not the case for Morocco and Mars and that correction was not done. 

      \subsubsection{ “Drag partitioning” correction. (Data of Morocco)}. 
The efficiency with which drag is partitioned between erodible and non erodible soils is
expressed as an increase in the threshold friction speed for saltation $u_t^*$ that depends on two
parameters $z_0 (m)$ (roughness length for momentum transfer) and $z_{0s} (m)$ -‘smooth’ roughness  by Marticorena and Bergametti (1995),  \cite{Marticorena&Bergametti_95}.
 
 As a first step, these values were taken as $z_0=10^{-3}m$ , and $z_{0s}=5*10^{-6}m$. 
This corresponds to particle beds of particle mean size of $D=150 \mu m$ (size of the particles composing a quiesent bed in Morocco desert).

 “Drag partitioning” correction can be calculated as 

 $ {u_t}^*_{correction-roughness}=1-(\ln(z_{0}/z_{0s})/\ln(0.35*((0.1/z_{0s})^{0.8}))$  by   \cite{Marticorena&Bergametti_95}.
And we have
$ {u_t}^*_{correction-roughness}\approx 0,28$ .
 This value   does  not consistent to the  measurements of Sensit. 
 
 To obtain a more acceptable estimate, the difference between wind friction threshold measured by Sensit  and wind friction threshold
  simulated without correction ($ {u_t}^*_{correction-roughness}$=1) was taken as “drag partitioning” correction:
  
 $ {u_t}^*_{correction-roughness}= 1-(\ln(z_{0}/z_{0s})/\ln(0.35*((0.1/z_{0s})^{0.8})) \approx 0,999$ 
  
 where $z_{0s} \approx 10^{-6} \mu m$ (bed particle size).
 
 $z_{0} \approx 1sm$
 
 $ {u_t}^*_{correction-roughness}=1$ means from physical point of view that mobilization process takes place at smooth roughness length.
 
 That  demonstrates how to estimated an unknown $z_0$ value using  measured Sensit data and simulated (non corrected 
 for roughness) wind friction threshold value.
 In the case of known value of $z_0$ it is possible also to estimate $z_{0s}$.
 
\subsubsection{“Moisture inhibition” correction“}

The increase in threshold friction velocity for saltation  $u_t^*$ due to soil water is represented by
the parametrization of Fecan \cite{FMB}. The capillary force is allowed to suppress dust deflation when the near-surface
soil gravimetric water content $w$ exceeds a threshold gravimetric water content $w_t$ \cite{Zender}.
 
 if $w > w_t$
 
$ {u_t}^*_{correction-moisture}=\sqrt{1.0+1.21(100*(w-w_t))^{0.68}}$

$w_t=mss_{frc-cly}*(0.17+0.14mss_{frc-cly})$  by Fecan \cite{FMB} 

or (as an assumption) it is possible  to remove the factor of $mss_{frc_cly}$ from $w_t$ to improve large scale behavior:
           
$w_t=0.17mss_{frc-cly}+0.14mss_{frc-cly}$

$ w = VWC\rho_{H2O}/ \rho_{dry}$- gravimetric water content 

$\rho_{H2O}=1000.0  (kg/m^3)$ - density of liquid water;

 $\rho_{dry}=\rho_{ParticlesSoil}*(1.0-WSWC)$
 
$ \rho_{ParticlesSoil}=2650.0 (kg/m^3)$ - dry density of soil particles (excluding pores);

 $VSWC$ - saturated volumetric water content;
 
 $VWC$ -  volumetric water content.
 The near-surface soil gravimetric water content $w$ in Morocco data did not exceed a threshold gravimetric water content $w_t$ and the correction of $u_t^*$ 
 for moisture content was not done. This correction was not applied for Mars data too.

\subsection{Horizontal flux estimation}
The horizontal flux of sand was estimated according to the theory of White, 1979 \cite{White}. 
\begin{equation}
\label{eq1}
Q_s=2.61 \rho (u_t^*/u^*)^3)(1-u_t^*/u^*)(1+u_t^*/u^*)^2/g
\end{equation}
where: $u_t^*$ - is the  threshold of wind friction, $u^*$ - is the wind friction, $g$ - is the constant of gravity, $\rho$ - is the density of air. 

 \subsection{Threshold wind friction and horizontal flux estimation taking into account the atmospheric electric field}
 
The horizontal flux of sand was estimated according to the theory of White, 1979 \cite{White}, the effect of electric forces following the ideas
proposed by Shao and Lu [2000] and formulation proposed by Jasper F. Kok 1 and Nilton O. Renno 2,\cite{Kok}. 
\begin{equation}
\label{eq1}
Q_s=2.61 \rho (u_{t_ef}^*/u^*)^3)(1-u_{t_ef}^*/u^*)(1+u_{t_ef}^*/u^*)^2/g
\end{equation}
where: $u^*$ - is the wind friction saltation, $g$ - is the constant of gravity, $\rho$ - is the density of air,
$u_{t_ef}^*$ - is the  threshold of wind friction, estimated according to \cite{Kok}:

$u_{t_ef}^=\sqrt{(A_n/\rho)(\rho_{particle}gd+(6G\beta/d\pi)-8.22\epsilon_{0}E^2/(c_s))}$ 

 where $A = 0.0123$ is a dimensionless parameter that scales the aerodynamic forces; $G$- is a geometric parameter that depends on the bed stacking and is of order 1 (Shao and Lu,
2000) \cite{Shao}, $\rho_{particle}$ - is the density of elevated particles, $E$ -  is the total electric field, $d$ - is the diameter of saltating particles, $\beta$ - is an 
empirical constant that scales the interparticle force and is on the order of $10^{-5}$ kg/s (Shao and Lu, 2000) \cite{Shao}, $\epsilon_{0}$ -is
the dielectric constant,  $c_s$  is a parameter which is equal to 1 for soils composed of perfectly spherical particles
and $0 < c_ s < 1$ for real soils composed of non-spherical particles.

\subsection{Data used for model validation on the Earth}

 During the 2013-2014 field  campaigns in Morocco two meteorological stations (equipped with the instrumentation described in details by Esposito et al., 2016 \cite{Esposito}) were deployed.  
Synchronized measurements comparable to those that are supposed to be acquired on Mars by the instrument DREAMS  
were done:  wind speed and direction, atmospheric pressure, solar radiation, air and soil humidity, air temperature, electric field and wind 
friction velocity.  The surface roughness length  (which  supposed to be estimated for Mars by using orbital photo and size distribution method \cite{Hebrard}) was measured  experimentally for the ground nearby the station in Morocco. 
Wind measurements were collected using 2D and 3D anemometers. Wind friction speed (described by Bagnold et al., 1941 \cite{Bagnold}) was recorded by using the measurements of
3D anemometer and applying the method of eddy correlations and Monin-Obukhov theory.
The vertical component of measured wind speed was negligibly small and that is why the approach of Zender \cite{Zender}, when the vertical wind component can be neglected, was used by us. 
Impact sensors and sand catchers of the instrumentation for real time aeolian sand transport rates measurement (the instrumentation "SENSIT") were used for sand flux registration.    
Data were taken every second. 
We compared simulated and measured  wind reference, wind friction, wind friction threshold and horizontal sand flux.
Two sets of experimental of data registered on 1) $08/08/2014$ and 2) $29/07/2014$ are represented in the article as 
 examples of typically "good" and typically "bad" conformity between the experiment and the model. 
 
 \subsection{Adaption of the model to the Mars environment}
 
The following  parameters: gravity, planet diameter, planet axis inclination, rotation velocity, distance from the Sun, albedo, atmospheric composition and density  
were changed. At the moment, coefficients of Monin-Obukhov stability correction functions are defined for Earth in the model. The surface morphology and surface layer heat exchange model 
are supposed similar to ones Morocco desert and that is why the surface type and properties were chosen similar to Morocco. Surface roughness length was taken equal to 1sm  according  to estimation  method of Hebrard \cite{Hebrard}. 
The heat flow was modified for Mars: dust and $CO_2$ infra-red radiation were included in it. Ice reflection is not included yet in the code. 
 The estimation of wind friction thrashold and sand/dust fluxes was changed for Mars according to the 
 works of Mulholland \cite{Mulholland}
  Considering that electrical dust raising mechanism is supposed to be important on Mars, the possibility to take into account the influence of atmospheric 
 electric field on simulated values was included in the model. 
\subsection{ Simulation of wind friction velocity  and threshold wind friction velocity on Mars.}

The methods of wind friction calculation for Mars are described in details by Mulholland \cite{Mulholland}. If lifting by wind stress takes place, the approach of Newman \cite{Newman} is used. 
 Newman assumes that saltation of sand particles is essential for lifting of micron-sized dust particles, and therefore the minimal 
 lifting threshold for an optimal size particle is calculated.  This threshold drag velocity is given by semi-empirical 
 formula  \cite{GrIv} that is adopted for Mars in \cite{Mulholland}.
 This semi-empirical wind friction threshold depends on atmospheric density, particle density and particle diameter, and the interparticle
cohesion parameter.  Several problems existe with this formulation. For example, the interparticle cohesion parameter has not been measured for Mars and the same values are commonly used for both 
Earth and Mars (Greeley and Iversen, 1985 \cite{GrIv}). 
To allow any dust lifting  on Mars at all it was necessary to use a value towards the lower end of allowed values; otherwise predicted thresholds were too high \cite{Mulholland}.
 It was found that extremely small (and probably unrealistic) values of the interparticle cohesion parameter were necessary to allow lifting arising \cite{Mulholland}.
 The particles can be lifted directly off the surface by the wind stress, and the minumal wind stress requared is called fluid threshold. 
The impact threshold is the lower of the two, since the 
contribution of the saltation flux enhances lifting, but on Earth the difference
between the two thresholds is not thought to be very great and their ration is about 0.8 \cite{Kok2}.
Recent numerical studies(Almeida et al., 2008 \cite{Almeida2}, Claudin and Andreotti, 2006 \cite{Claudin}, Kok, 2010a \cite{Kok2}) have found that the situation on Mars may be quite different. It
has been estimated that the impact threshold may be less than 50\% of the fluid threshold: 0.48\% according to Almeida et al.\cite{Almeida2} ; 0.3\% according to Claudin and Andreotti\cite{Claudin} and
0.1 according to Kok \cite{Kok2}.
 
 A  simpler estimation of threshold wind friction, which avoids the iterative calculation required previously, was found using the approach of Shao and Lu \cite{ShaoLu}.  In their model of
saltation, the interparticle cohesion is assumed to be a function of particle size and does not explicitly appear in the formulae. It nonetheless produces a similar size 
dependence to the Greeley and Iversen  method \cite{GrIv}, with a minimum threshold for sand-sized
particles, in agreement with experimental studies of Greeley and Iversen \cite{GrIv} and  \cite{Mulholland}.  This formula can be differentiated with respect to particle diameter to find the minimum threshold velocity. 
The threshold velocity  depends on on near-surface atmospheric density in this formula.  The previous method of calculating the threshold had a more complex functional
dependence on air density, but actually produced quite little deviation from the constant threshold approach, so this simpler method arguably sacrifices little accuracy.
Indeed, Newman et al.\cite{New}  reverted to such a method, for experiments using altered orbital parameters. The method of Shao and Lu \cite{ShaoLu} requires no unrealistic assumptions 
of parameter values. 
However, according to \cite{Mulholland}, thresholds calculated using the method of Shao and Lu \cite{ShaoLu} still proved to be too
high for lifting by model surface winds, except in very rare cases. 
 
 The Shao and Lu   calculation \cite{ShaoLu} is for the fluid threshold and will overestimate the difficulty of lifting — in fact, surface wind need only exceed the fluid threshold 
 briefly for saltation to begin, after which lifting can continue so long as the wind stays above the lower impact threshold. 
 It seems that a considerable hysteresis effect may be at work for Martian dust lifting.

 The lifting threshold formulation used by global and mesoscale modeling is an approximation to both the issues mentioned above, applied in such a way as to
remain relatively simple whilst allowing some variability.  Both sub-gridscale gustiness and the effect of a low impact threshold should lead to more dust
lifting. Mulholland \cite{Mulholland} combined both these considerations (fluid and impact threshold for Mars) into a single reductive scaling factor,
applied to the fluid threshold. The lifting threshold was lowered until an appropriate amount of dust lifting became possible. The value of this scaling factor depends somewhat on resolution, but at T31 a value of 0.7 was found to be appropriate. 
This is obviously a major simplification of what are undoubtedly important effects in Martian dust lifting, but at present it seems the most logical
approach to take for a GCM lifting scheme, until further information is available \cite{Mulholland}.

Eddy simulation (LES) studies promise to tell us more about the sub-gridscale surface vertical wind variations, but high resolution for horizontal winds is also needs. 

\subsection{Data used for test on Mars}

The data of Viking Lander for days when dust storm was recorded were used. 
These days were chosen, because, in such conditions, the friction of the wind should be greater than 
the threshold wind friction. And accordingly, it is expected that the simulated wind friction should be above the threshold, which is some kind of verification of the model.

 \section{Test of the model on Morocco data. Results and Discussion.}
  
 
Zender's model is valid in the conditions of the logarithmic wind profile. 
Although only one wind level was used for the presented simulation, wind speeds  were registered at 0.5, 1.41, 4.5, 5, 7 and 10 m of height. And it was a chance to check whether the wind profile corresponds 
to a logarithmic one. Wind speed gradient was computed by averaging the measurements at each height during 
each 6 hours period of data registration (from 00:00 to 6:00, from 6:00 to 12:00, from 12:00 to 18:00 and from 18:00 to 24:00) as well as during all 24 hours of each runtime. 
The best correspondence of wind profile to the logarithmic shape was during daytime hours (from 09:00 to 15:00).

The correspondences of wind profile to the logarithmic shape for both data sets are shown on the Figure 1.  Pearson coefficients of correlation between the mean (24 hours) 
experimental data and its logarithmic curve fit were $0.994$ and $0.939$ for $08/08/2014$ and $29/07/2014$ respectively.   
 
 \begin{figure}[h!]
\centering
\includegraphics[width=1.2\textwidth]{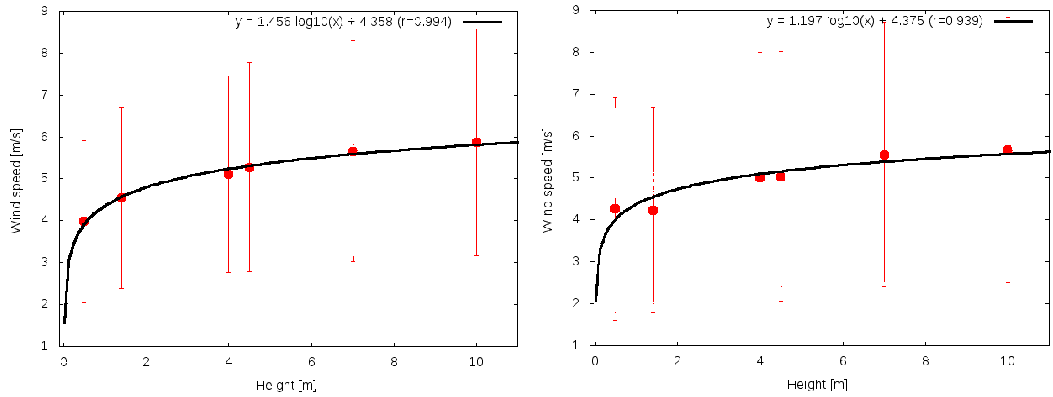}
\caption{Y - wind speed (m/s) averaged during 24-hour interval, X - height of measurement (m)}
\label{fig1}
\end{figure}

 \subsection{Сomparison of simulated and measured reference wind velocity}
Simulated and measured wind speeds at the height of $10m$ are represented on the Figure 2. Pearson coefficient of correlation between the experimental data and the simulated curve
was $0,762$ for $08/08/2014$ and  $0,991$ for $29/07/2014$ respectively. 

\begin{figure}[h!]
\centering
\includegraphics[width=1.4\textwidth]{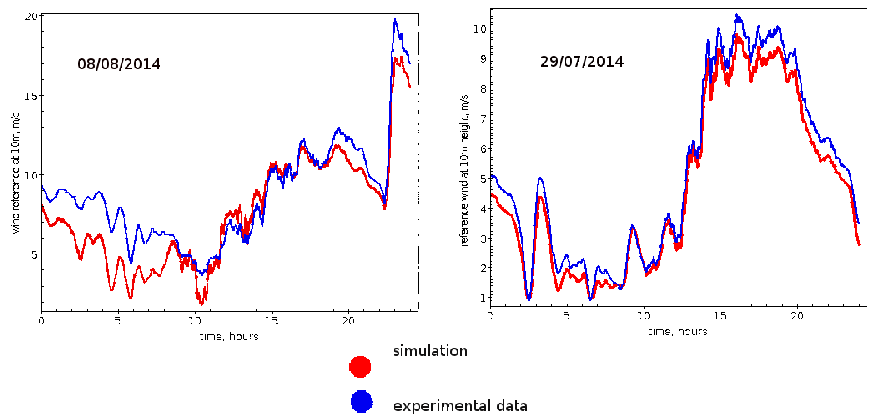}
\caption{ 08/08/2014; X - time (hours), Y - reference wind  speed at 10m (m/s)}
\label{fig2}
\end{figure}

 \subsection{Сomparison of simulated and measured wind friction velocity}
As described above,  wind friction velocity was modeled by three different methods.
The simulated wind friction velocities were compared with the experimental data. The results are represented on the Figure 3.  

\begin{figure}[h!]
\centering
\includegraphics[width=1.2\textwidth]{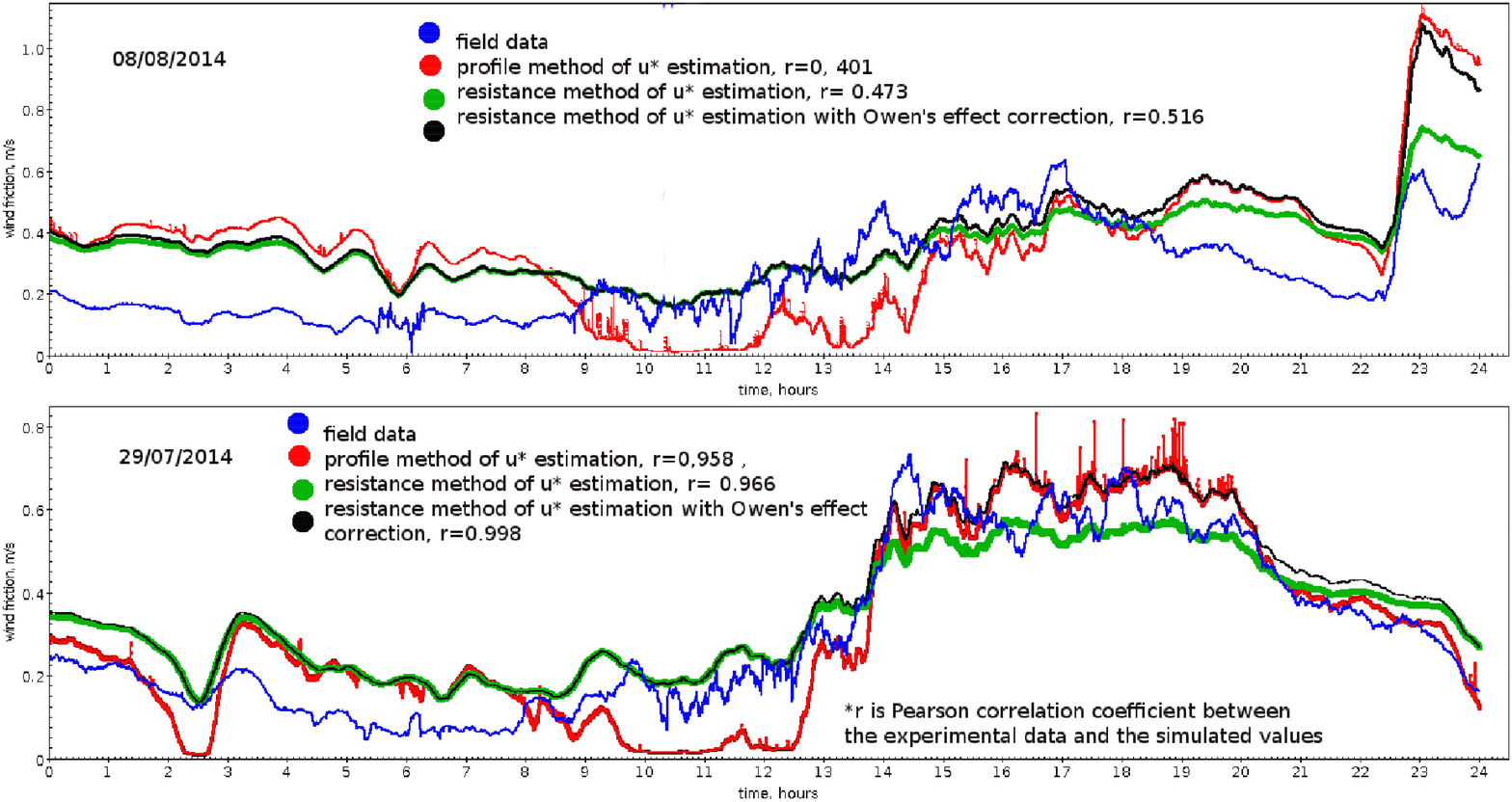}
\caption{ Wind friction speed.  X - time (hours), Y - wind friction speed (m/s)}
\label{fig3}
\end{figure}
 
 The parameters of statistical correlation analysis between the experimental data and the modeled wind friction are the following:
 
 $08/08/2014$   Wind friction deposition: model versus data

  Mean:                  0.4414                0.3477

  Std Dev:               3.3768                0.2228
  
  Sum:                   38132.0000            30034.0977
  
  Sum of Squares:       1.00195e+06            14731.4610
 
  Pearson Correlation Coefficient:  r = 0.4011
  
$08/08/2014$   Wind friction mobilization: model versus data                             
                 
  Mean:             0.4414            0.3477
    
  Std Dev:            3.3768              0.2228
  
  Sum:                38132.0000            30034.0977
  
  Sum of Squares:     1.00195e+06            14731.4610
 
  Pearson Correlation Coefficient:  r = 0.4730
 
$08/08/2014$  Wind friction saltation: model versus data     
                 
  Mean:          0.4414              0.3477
   
  Std Dev:       3.3768               0.2228
  
  Sum:            38132.0000            30034.0977
  
  Sum Sq.:       1.00195e+06            14731.4610
  
  Pearson Correlation Coefficient:  r = 0.5165
  
  $29/07/2014$  Wind friction deposition: model versus data   
            
  Mean:          0.3041                0.3422
 
  Std Dev:       0.1979                0.1377
  
  Sum:          26269.4073            29563.7894
  
  Sum Sq.:      11370.8089            11756.1182

  Pearson Correlation Coefficient:  r = 0.9584
 
  $29/07/2014$ Wind friction mobilization: model versus data
 
  Mean:          0.3041                0.3760
  
  Std Dev:       0.1979                0.1803
  
  Sum:         26269.4073            32483.9862
  
  Sum Sq.:      1370.8089            15023.7510

  Pearson Correlation Coefficient:  r = 0.9659
  
  $29/07/2014$ Wind friction saltation: model versus data

  Mean:          5.2030                4.7176
  
  Std Dev:       2.9706                2.8268
  
  Sum:          449528.2099           407590.7247
  
  Sum Sq.:      3.10130e+06           2.61325e+06

  Pearson Correlation Coefficient:  r = 0.9983
  
The results show a good accordance between modeled and  measured wind friction velocities. The best matching corresponds to the time of near neutral or 
slightly unstable atmospheric conditions. This results is in agreement with Jacovides et al., (1992) \cite{Jacovides}.

  Equations based on the Monin-Obukhov theory were used to account for stability effects through parameterization that is different for  profile and  aerodynamic resistance methods. 
  Aerodynamic resistance method parameterizes the kinematic fluxes in terms of mean quantities which are evaluated as follows: 
 1) the surface kinematic fluxes of momentum and the latent/sensible heat are defined from the Monin-Obukhov theory applied to the surface layer (constant flux layer), as described by Brutsaert et al., 1982  \cite{Brut} and Arya et al., 1988 \cite{Arya}; 
 2) the aerodynamic and surface resistances are calculated from the measured meteorological data. 
The profile method specifies surface temperature rather than solving the energy balance equation like in the aerodynamic resistance method.
   
The best correspondence between the measured and the simulated wind friction velocity was found for the aerodynamic resistance method with the Owen's correction 
(for $u^*_{saltation}$). 

\subsection{Сomparison of simulated and measured threshold wind friction velocity} 
  Simulated and  measured threshold wind friction velocities are represented on the figure 4.  
  Threshold friction velocity was defined experimentally as the wind friction velocity at which saltation initiates: the sensor device "SENSIT" activates.
  
 $27/09/2014$ - the modeled threshold wind friction changed from 0.218 to 0.225 m/s while the experimental threshold wind friction was about 0.256 m/s.
 
 $08/08/2014$ - the modeled threshold wind friction changed from 0.218 to 0.224 m/s while the experimental threshold was about 0.250 m/s, figures 4.
 
 \begin{figure}[h!]
\centering
\includegraphics[width=1.0\textwidth]{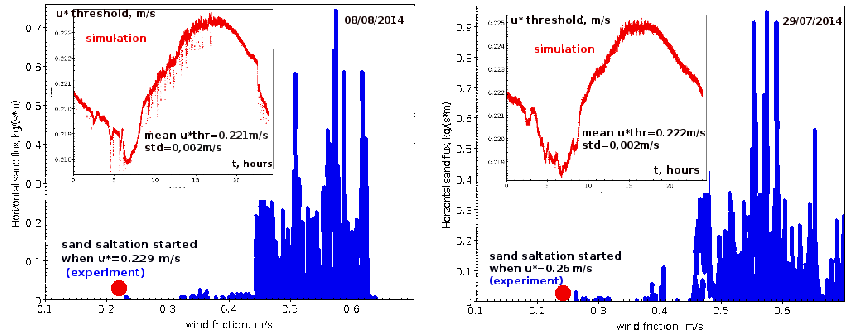}
\caption{Threshold wind friction. Small figure, red curve: axis Y- threshold wind friction speed, (m/s), axis Y time step,(hours). Lower figure, blue: experimental 
measurement of threshold wind friction speed. Axis X - threshold wind friction speed (m/s). Axis Y- horizontal flux of sand ($kg/(s*m)$). Saltation start is marked with a red dot. 
The wind friction speed at this point was taken as the experimental value of threshold wind speed.} 
\label{fig4}
\end{figure}

 $u^*_{saltation}$ was used to calculate the  wind friction speed.  We estimated the impact wind friction threshold assuming  
 that direct dust lifting does not take place for Morocco conditions and the fluid wind friction threshold exceeds the impact wind friction threshold. 
 
  The threshold wind friction changes dramatically with soil condition variations. Characteristics like surface roughness length and 
  mass of clay as well as moisture, organic matter and salt content noticeably affect the threshold wind friction. A small amount of 
  precipitation or an established crust can have an extreme effect on threshold velocity. All these factors are not explicitly 
  parameterized.  For example, the effects of  moisture, salt and organic matter content are not well documented.
  For example, when surface roughness length is changed from $1sm$ to $3sm$,
   wind friction threshold increases twice. 

Sandblasting and disaggregation of small clay and silt-sized particles from the surface and the larger particles during saltation, affect the wind friction threshold and 
are strongly sensitive to the size distribution of saltating particles (Shao et al., 1996) \cite{Shao}; (Grini et al., 2002) \cite{Grini}.

The model assumes that the saltation process is initiated whenever $u^* > u^*_t(D_0)$, where $D_0$ is the diameter of the 
"optimal particle" with minimum $u^*_t)$. 
Therefore, the estimated wind friction threshold is probably underestimated. 
 
Nevertheless, a good correspondence between measured and simulated value of threshold wind friction was found. 

 \subsection{Сomparison of simulated and measured horizontal sand flux}
 The simulated and the registered horizontal flux of sand for both data sets ($08/08/2014$ and $27/09/2014$) are plotted on the figure 5.

  \begin{figure}[h!]
\centering
\includegraphics[width=1.0\textwidth]{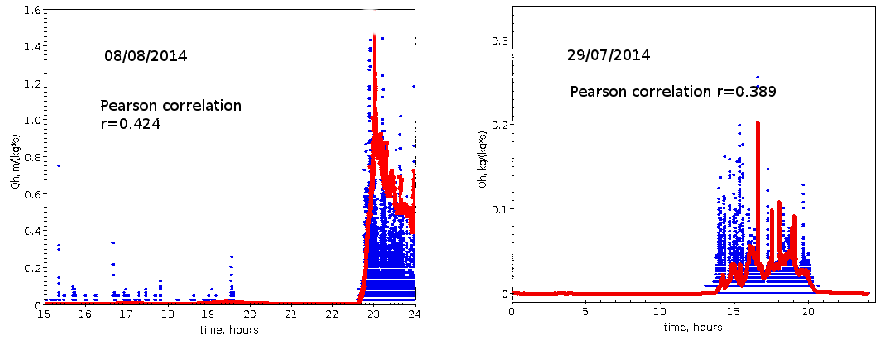}
\caption{Horizontal flux of sand. Red - simulation, blue - experimental measurements. 
axis X - time, (hours), axis Y - horizontal flux of sand($kg/(s*m)$).}
\label{fig5}
\end{figure}
The results of reference wind, wind friction and wind friction threshold simulation agree well with the field measurement, but  
the conformity between the modeled and the experimental horizontal sand fluxes is not so good. 

The Pearson correlation between measured and simulated data was 0,424 for $08/08/2014$ and 
0,389 for $29/07/2014$. 

The following factors could be the possible causes of this discrepancy: 1) experimental errors and inaccuracies in flux measuring, for example, 
precise measurement of large sand fluxes is difficult due to the construction of the recording device; 
2) the method of White \cite{White} for flux estimation is imperfect; 
3)  electric field influence on dust lifting was not taken into account. 

Horizontal flux of sand estimated taking into account the influence of electric field on sand/dust lifting 
was  1.5 - 2 times greater than horizontal flux of sand estimated without electric field influence. 
Time profiles of simulated and measured fluxes are presented on the figure 6.  
 
  \begin{figure}[h!]
\centering
\includegraphics[width=1.2\textwidth]{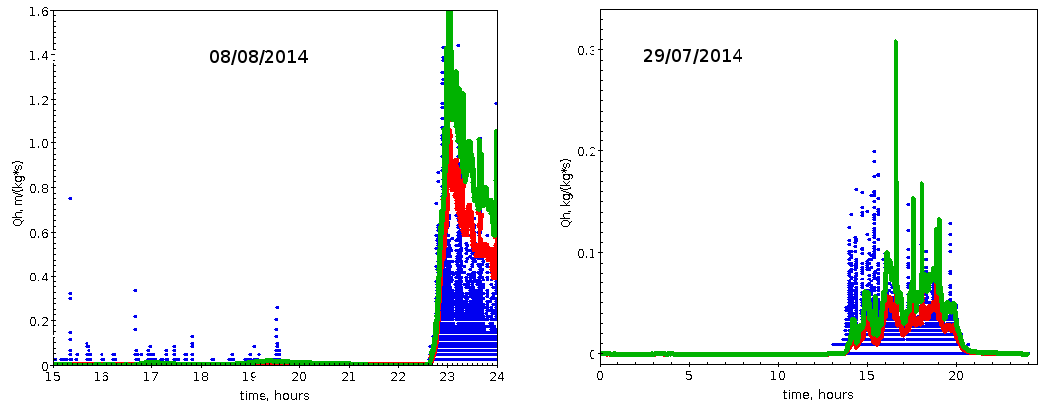}
\caption{{Horizontal flux of sand. Red - simulation without electric field influence included, blue - experimental measurements, green -simulation with 
electric field influence included. 
Axis X - time, (hours), axis Y - horizontal flux of sand($kg/(s*m)$).}} 
\label{fig6}
\end{figure}

\subsection{Possible sources of error.}
Testing the model on the data measured in the desert of Morocco, 
where climate and soil parameters are similar to the Martian conditions, gave a good result.
Simulation errors result may arise from the uncertainty of input data measurements.  The assumptions and limitations of the model influence the output results. 
The values of the empirical coefficient used can affect the results, for example, the Karman universal constant ranges from 0.35 to 0.42. The stability correction 
function in Monin Obukhov equations is empirical.

\section{Test of the model on Viking Lander data. Results and Discussion.}
The model can run on available  Viking Lander, Pathfinder or Phoenix  data. Unfortunately,  simulated wind friction, wind friction threshold, horizontal sand flux and vertical dust flux  cannot be verified experimentally  due to the lack of appropriate measurements on Mars now. 
But it is expected that future Mars rovers will provide these parameters.
This article presents two modeling output parameters for Mars: wind friction and wind friction threshold. 
It is these quantities that were chosen since at the moment these values are  critical points of dust lifting modeling on Mars. 

Four different methods were used for wind friction thresholds estimation:
1) by formulation of Newman et al. \cite{New} (impact wind friction threshold);
2) by formulation of Shao and Lu \cite{ShaoLu} (fluid wind friction threshold);
3) by formulation of Mulholland \cite{Mulholland} (corrected fluid wind friction threshold);
4) by  formulation of Kok, \cite{Kok} (electrical wind friction threshold). 
  
 The results are following (figures 7 and 8):
\begin{itemize} 
 \item1) impact wind friction threshold for saltation - 1,17 m/s ,
 \item2) fluid wind friction threshold - 1,61 m/s ,
 \item3) corrected fluid wind friction threshold - 1,16 m/s ,
 \item4) "electrical" wind friction threshold  - from 0,32 m/s to 0,63 m/s depending on the magnitude of total elecric field that varied in simulations 
 from 10 to 50 kV/m.     
\end{itemize} 

 The simulated  wind friction changed from  0,1m/s to 0,63 m/s (figure 8). 

 The thresholds calculated without taking into account the influence of the electric field turned out to be greater than the calculated wind friction, which makes the rise of dust with such thresholds impossible. Given that the input for the simulation was measured during a dust storm, the estimated wind friction should be greater than the threshold value. Therefore, thresholds simulated by the Kok method with a field value from 40 to 50 are plausible.
The simulation results are of course preliminary and the model needs refinement, nevertheless, such an approach  can be promising.
 \begin{figure}[h!]
\centering
\includegraphics[width=1.0\textwidth]{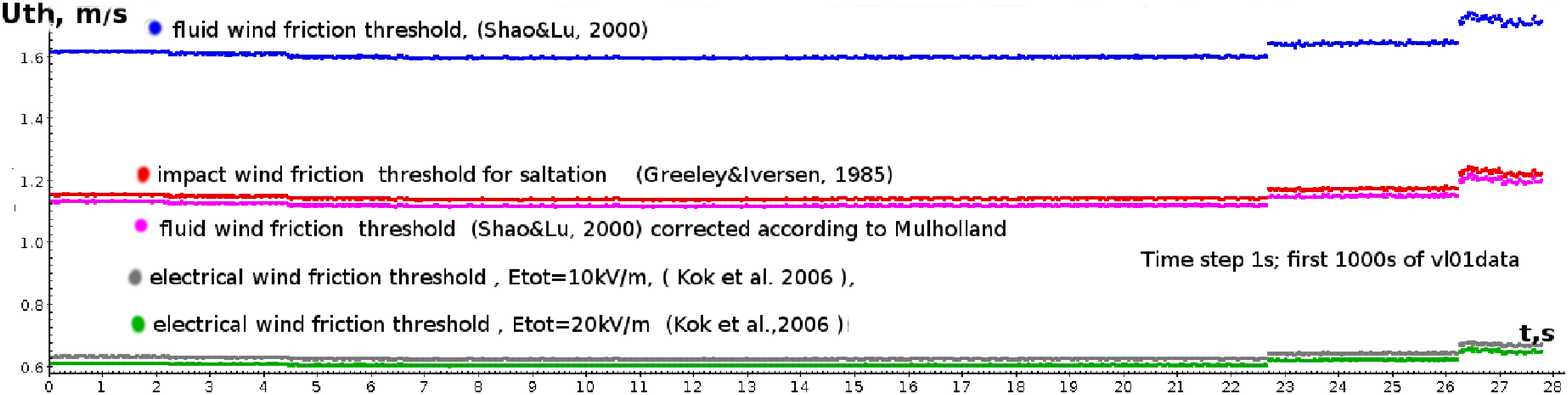}
\caption{Treshold wind friction. Simulation with local wind erosion and dust deposition program for Mars. 
Input data - Viking Lander 1;Vl01.
 Axis X:  time, s. Axis Y: red line- simulated impact threshold wind friction saltation, (m/s); green lines - simulated threshold wind friction with electric field influence included, (m/s); 
 blue - simulated wind friction, (m/s).} 
\label{fig7}
\end{figure}

  \begin{figure}[h!]
\centering
\includegraphics[width=1.0\textwidth]{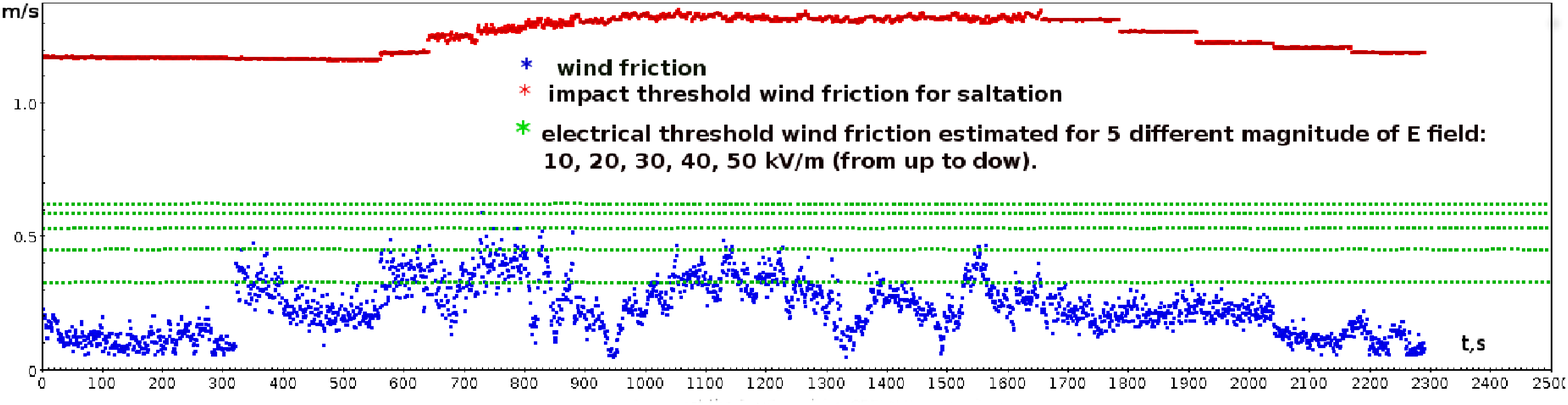}
\caption{Wind friction and threshold wind friction. Simulation with local wind erosion and dust deposition program for Mars. Input for the simulation Viking lander data.Vl015.
 Axis X:  time, s. Axis Y: red line- simulated impact threshold wind friction saltation, (m/s); green lines - simulated threshold wind friction with electric field influence included, (m/s); 
 blue - simulated wind friction, (m/s).} 
\label{fig8}
\end{figure}

 \section{Conclusions}
 
The stated motivation of this work was to prepare the local wind erosion and dust deposition model in a way that will eventually enable its use for predicting dust emission/lifting 
rates on Mars (for specific use in GCMs).   The model implementation was planned in close conjunction with the upcoming DREAMS suite of instruments on the ExoMars 2020 surface platform.   
 First, the adaptation  of the local wind erosion and dust deposition model to the environment of the Merzouga desert was fulfilled. 
 The influence of the atmospheric electric field on dust lifting was included in the model. 
Morocco field-campaign 2014 data were chosen for simulation test. 
The simulation was performed by using the same input parameters that supposed to be supplied by DREAMS
 mission. The calculated wind reference, wind friction, threshold wind friction and horizontal flux of sand were in agreement with measured data.  
 After the test in Earth conditions, the model was adapted  for Mars environment. 
Impact threshold wind friction and electrical threshold wind friction were modeled using the Viking Lander data as available inputs aiming to demonstrate that the model is applicable for Mars. 
The  new local wind erosion and dust deposition model may serve a source of vertical dust flux as well as others local input parameters for the MGCM, because it 
is supposed to be more accurate than global, mesoscale simulations and Eddy simulation for local horizontal winds. 
\section{Bibliography}


\end{document}